\documentclass[pre,preprint,tightenlines,showpacs,nofootinbib,showpacs]{revtex4}

\usepackage{graphicx,bm,amsfonts}
\usepackage[tbtags]{amsmath}

\newcommand{\ij}{i\kern -0.08em j}
\newcommand{\half}{{\textstyle\frac{1}{2}}}

\def\im{\mathop{\rm Im}\nolimits}
\def\re{\mathop{\rm Re}\nolimits}

\def\beq{\begin{equation}}
\def\eeq{\end{equation}}
\def\eeql#1{\label{#1} \end{equation}}
\newcommand{\ket}[1]{\mathopen|#1\rangle}
\newcommand{\ua}{\uparrow}
\newcommand{\da}{\downarrow}
\newcommand{\To}{\rightarrow}

\newcommand{\pt}{\partial}

\newcommand{\Oc}{\mathcal{O}}
\newcommand{\op}{\omega_\mathrm{p}}

\DeclareMathOperator{\sgn}{\mathrm{sgn}}

\begin{document}

\title{Resonance eigenstates of the SQUID--qubit system}
\author{Alec \surname{Maassen van den Brink}}
\thanks{Presently at: Frontier Research System, RIKEN, Wako-shi, Saitama, 351-0198, Japan}
\email{alec@riken.jp}
\affiliation{D-Wave Systems Inc., 100-4401 Still Creek Drive, Burnaby, B.C., V5C 6G9 Canada}
\date{\today}

\begin{abstract}
We study the complex-valued resonance spectrum of a dc-SQUID coupled to a flux qubit, where the former is treated in the cubic and the latter in the two-level approximation. It is shown that this spectrum is well-defined and contains most of the relevant information on the escape process. Thus, the language of resonance states is precise and well-adapted to switching- (or trigger-) type qubit readout, and a worthwhile complement to the various descriptions of continuous qubit measurement. Initial progress is analytic, but nonperturbative numerical methods have been formulated and should soon yield accurate results for all parameter values.
\end{abstract}

%measurement theory
\maketitle

\section{Introduction}

Parallel to, and stimulated by, the remarkable development of experimental quantum computing, there has been considerable theoretical attention for the problem of measuring a single two-level system (qubit, pseudospin). There is considerable variety both in the studied detector candidates (quantum point contacts \cite{gurvitz,korotkov,goan,KA}, normal~\cite{MSS,cot} and superconducting~\cite{cottet} single-electron transistors in various regimes) and in the employed theoretical apparatus (correlation functions and diagrammatic methods~\cite{KA,cot}, generalized Master equations~\cite{gurvitz,MSS}, quantum trajectories~\cite{goan}, and Bayesian formalism~\cite{korotkov}). However, the theoretical emphasis seems to be on \emph{continuous} readout while experimentally, trigger-type detectors are dominant, at least until recently. In the latter, exemplified by switching-current measurement in a dc-SQUID, the readout is a single event instead of a continuous process, making the theoretical description rather different. Yet, many physical questions, such as how to characterize detector efficiency, and what exactly is measured when reading out a superposition state in the presence of a non-negligible qubit Hamiltonian, are similar. It is our purpose to study these issues, using a language especially suited to trigger-type readout.

The model consists of a dc-SQUID coupled to a flux qubit, described by the Hamiltonian
\beq
  H=-\frac{2e^2}{C}\frac{d^2}{d\phi^2}+\alpha\phi-\beta\phi^3
  -\half(\epsilon\sigma_z{+}\Delta\sigma_x)+\lambda\sigma_z\phi\;.
\eeql{H}
The first three terms represent a small-inductance dc-SQUID for a near-critical external current bias, where the metastable wells of the tilted-washboard potential are widely spaced relative to their width and depth, so that for the escape process one can focus on a single such well. Clearly, one needs $C>0$ for the capacitance in order to have a well-behaved mass term; taking $\beta>0$ establishes $\phi\To\infty$ as the unstable direction, upon which one also needs $\alpha>0$ to have a potential well near $\phi=0$. We also take $\Delta>0$ for the qubit tunneling amplitude, and $\lambda>0$ for the SQUID--qubit coupling. The latter breaks the equivalence of $\sigma_z=\pm1$, so it is better to let $\epsilon$ have either sign. As written, the model~(\ref{H}) has six parameters,\footnote{Achieved in part by taking the inflection point of the potential as the origin of~$\phi$, which eliminates the $\phi^0$ and $\phi^2$ terms.} but it will be shown in Sec.~\ref{prelim} that there are only four nontrivial dimensionless ones.

A remark is in place concerning the choice of coupling term $\lambda\sigma_z\phi$: while $\sigma_z$ represents the qubit flux in the two-state approximation, $\phi$ in (\ref{H}) is the overall phase difference in the dc\nobreakdash-SQUID (conventionally, the average of the two junction phases), not the latter's loop current (related to the difference of these phases), which in fact gets eliminated when modeling a small-inductance SQUID with a single degree of freedom. Thus, the term does not simply represent inductive flux--flux coupling. One can, however, reason as follows: while $\beta$ is fixed for a given SQUID, $\alpha$ is proportional to the (small) difference between the bias and effective critical currents, where the latter is dependent on the external flux applied to the SQUID; in its turn, this flux may effectively contain a component inductively coupled into the SQUID loop by the qubit circuit. If the external influence on $\alpha$ is linearized, this mechanism is reflected by our choice $(\alpha+\lambda\sigma_z)\phi$ in~(\ref{H}). Foremost, though, the model~(\ref{H}) is meant to be physically motivated and well-defined for all parameter values; note that a metastable and essentially irreversible trigger, coupled to a quantum two-level system, may also be a prototype for, say, particle detectors. For an accurate description of a particular (strongly coupled) SQUID--qubit system, it is advisable to do a more systematic derivation, starting from a multi-loop, multi-junction Hamiltonian.

The SQUID potential $U(\phi)=\alpha\phi-\beta\phi^3$ allows for outgoing states which never return to the well, and a well-defined escape problem results without need for a heat bath or other model of dissipation.\footnote{Accurately modeling a given experimental situation may still require the inclusion of such a heat bath, of course. In particular, a switching-current distribution wider than predicted by (\ref{H}) may indicate the presence of additional noise sources.} In this setting, one expects the existence of exact \emph{resonance states} or \emph{quasinormal modes}
\beq
  \ket{\Psi(\phi,t)} =
  \begin{pmatrix} \psi_\ua(\phi,t) \\ \psi_\da(\phi,t) \end{pmatrix} =
  \ket{\Psi(\phi)}\,e^{-i\omega t}\;,
\eeql{QNM}
where $\omega$ is \emph{complex} with $\im\omega<0$. Thus, the exponential in (\ref{QNM}) accounts for both an ``energy'' $\re\omega$ and a decay rate $\lvert\im\omega\rvert$ (for the amplitude; the probability decay rate is twice this). The notation $\ket{\Psi}$ is merely shorthand for the two-component wave function; in particular, there is no implication that $\ket{\Psi}$ is normalizable, although some linear-space structure will be introduced in Sec.~\ref{analytics}. This also means that, while $\ket{\Psi}$ and $\omega$ should follow from the two-component eigenvalue equation
\beq
  H\ket{\Psi}=\omega\ket{\Psi}\;,
\eeql{SE}
the latter's familiarity is in part deceptive. While we thus deviate from textbook Schr\"odinger theory, exact resonances are known for, e.g., Fokker--Planck (generalized diffusion) problems~\cite{FP}, in cavity optics, and in linearized black-hole gravity~\cite{RMP}. If one has an effective way to find these resonances also for~(\ref{SE}), the above defines a framework for switching-type (discontinuous) quantum measurement of perhaps unprecedented rigour and precision, for arbitrarily large escape rates. An elementary introduction to quasinormal modes of the Schr\"odinger equation is given in the appendix.

The challenge becomes to calculate the resonance spectrum $\{\omega_j(C,\alpha,\beta,\epsilon,\Delta,\lambda)\}$, in a setting where the standard Schr\"odinger toolbox is called into doubt. Based on the large-$\phi$ asymptotics, it is suggested to study (\ref{SE}) not along the real $\phi$-axis, but along a contour in the complex plane for which the boundary values are more readily implemented. In Sec.~\ref{free}, this is taken up for the noninteracting
\beq
  H_\mathrm{S}=-\frac{2e^2}{C}\frac{d^2}{d\phi^2}+\alpha\phi-\beta\phi^3\;.
\eeql{H0}
This contour approach is related to, and hopefully a refinement of, the ``complex scaling'' method in~\cite{yaris}; at the least, results will be compared. Since the problem's instability/non-hermiticity is solely due to the SQUID, (\ref{H0})~should capture the essence of these aspects, upon which addition of the qubit presents only technical complications, addressed in Sec.~\ref{int-num}. (Of course, the qubit complicates the interpretation of the results.)

With the uncoupled problem (\ref{H0}) under control, Sec.~\ref{int} returns to the interacting Hamiltonian~(\ref{H}). In Secs.\ \ref{int-analytics} and~\ref{int-deg}, this is first of all done perturbatively; interestingly, the analysis applies a bit beyond the weak-coupling regime, in that a finite quantum non-demolition (QND) qubit--detector interaction is admissible. Subsequently, in Sec.~\ref{int-num}, a numerical scheme for the interacting case is formulated, the results of which should provide general and accurate information, and allow comparison with analytical results such as in~\cite{grabert}.

As alluded to above, this work is motivated by questions about the interpretation of switching-current measurement. The qubit will affect the escape process in the SQUID so the latter may be said to read out the former, but how exactly? A plausible scenario, for which there is some experimental support~\cite{NTT}, is the ``self-averaging'' one: for each qubit eigenstate~$k$, the SQUID sees the expectation of the qubit flux in that particular eigenstate, i.e., $\alpha\mapsto\alpha+{\langle\sigma_z\rangle}_k\lambda$. The underlying physical picture is that the influence of the qubit on the SQUID is to be averaged over many coherent oscillations in the former. However, one can also argue in favour of a ``weakest-link'' scenario~\cite{AS}, in which the SQUID predominantly escapes when the qubit is in the state $\sigma_z=-1$, facilitating said escape, i.e., $\alpha\mapsto\alpha-\lambda$ whenever the state $\sigma_z=-1$ has an appreciable probability of being occupied. While less prominent in the literature, this \emph{is} the behaviour one would expect of a readout SQUID coupled to a classical oscillator. At least for a linear detector, in Sec.~\ref{int-deg} these two scenarios are recognized as opposite limits of a more comprehensive description. Besides this practical relevance for SQUID-based flux measurement, the aim is to contribute to quantum measurement theory in general, and show how the same concepts surface in radically different readout methods.

\section{Free SQUID}
\label{free}

\subsection{Preliminaries and complex-contour formulation}
\label{prelim}

Consider $(H_\mathrm{S}-\omega)\psi(\phi)=0$, with $H_\mathrm{S}$ as in~(\ref{H0}). The full titled-washboard Hamiltonian has three parameters. Eliminating the overall energy scale, one would be left with two dimensionless parameters: one characterizing how close one is to critical bias (a purely classical property), and one counting the number of levels in each well (also involving the capacitance). In the cubic approximation, one always is infinitesimally close to critical bias, so one expects only the latter parameter to remain.\footnote{One can also consider the cubic problem as a harmonic oscillator with a cubic ``correction". Since all harmonic oscillators are created equal, only the magnitude of the cubic term is a nontrivial parameter.} Indeed, scaling $s=\kappa\phi$, one finds that for $\kappa=(\beta\gamma)^{1/5}$ (with $\gamma\equiv C/2e^2$) the Schr\"odinger equation takes the form
\beq
  -\psi''+(\tilde{\alpha}s-s^3-\tilde{\omega})\psi=0\;,
\eeql{scale}
where now of course $\psi''=d_s^2\psi$, and where the new parameters are related to the old ones as $\tilde{\alpha}=(\gamma^{2/5}/\beta^{3/5})\alpha$ and $\tilde{\omega}=(\gamma^{3/5}/\beta^{2/5})\omega$. While one could also have left the sole parameter in front of either the kinetic or the cubic term, the form (\ref{scale}) has two slight advantages: (A)~it still has the flexibility to tune the potential from metastable via critical to unstable, and (B)~the leading asymptotic behaviour for $|s|\To\infty$ is independent of $\tilde{\alpha}$, see around (\ref{psiR}) below. Since there is no need to refer to the unscaled problem any more except for comparing to experimental parameters, we will proceed using the form (\ref{scale}) and henceforth omit the tildes.

Next, let us study (\ref{scale}) asymptotically. The leading WKB behaviour is $\psi(s)\approx\exp\bigl[\pm\int^s\!du\,\sqrt{U(u)-\omega}\bigr]$. For each $\omega$, define $\psi_\mathrm{L}(s,\omega)$ and $\psi_\mathrm{R}(s,\omega)$, satisfying the ``left" ($s\To-\infty$) and ``right" ($s\To\infty$) boundary conditions, respectively, but in general not both. For $\psi_\mathrm{L}$, the potential $U(s)\To\infty$ presents a conventional wall, and one clearly has to take the bounded $\psi_\mathrm{L}(s)\sim e^{-\frac{2}{5}(-s)^{5/2}}$. For the ``right" solution $\re s\To\infty$, matters are more complicated, since neither solution
\beq
  \psi_\mathrm{R}(s)\sim
  \exp\biggl[\pm i\int^s\!\!du\,\sqrt{\omega-U(u)}\biggr]
\eeql{psiR}
dominates the other [i.e., the positive real $s$-axis is an \emph{anti-Stokes line} of~(\ref{scale})]. One now reasons as follows: since $\psi_\mathrm{R}(s,\omega)\propto G(s,u;\omega)$ for $s>u$ [see~(\ref{G}) below] and $G$ is the retarded Green function, $\psi_\mathrm{R}(s,\omega)$ should be bounded in the upper half $\omega$-plane. Hence, one should take the plus sign\footnote{This selection of the outgoing wave also follows by requiring that, for real $\omega$ and $s\To+\infty$ in (\ref{psiR}), one should have $J>0$ for the probability current $J=i(\psi\nabla\psi^*-\psi^*\nabla\psi)$.} in (\ref{psiR}), i.e., $\psi_\mathrm{R}(s)\sim e^{i\frac{2}{5}s^{5/2}}$. Further, one now finds that $\psi_\mathrm{R}(s)\sim e^{-\frac{2}{5}|s|^{5/2}}$ along the line $\arg s=\pi/5$, the \emph{Stokes} line in the upper-half $s$-plane closest to $s>0$. Hence, it is suggested that any numerical analysis proceed asymptotically along this line, for maximum stability and minimum oscillation of the solution. Because of the general analyticity properties of solutions to linear analytic ODEs, and the nearly ideally simple behaviour of $U(s)$ in the complex $s$-plane, the choice of contour $\mathcal{C}$ joining the ``left'' and ``right'' asymptotes is now flexible and can be based on numerical convenience. A hyperbola comes to mind, but two straight lines joining at the origin may be even simpler.

The next step is actually solving $H_\mathrm{S}\psi=\omega\psi$. It is stressed that the above WKB asymptotics were intended only for finding the contour~$\mathcal{C}$; the eigenvalues themselves are to be found numerically, with in principle arbitrary precision. At least two methods seem viable. (A)~Discretization: the continuous problem (\ref{scale}) is converted into a matrix problem, which is subsequently diagonalized either fully or partially. Care has to be taken when covering the infinite $s$-contour with a finite number of grid points. (B)~Searching for roots of the Wronskian
\beq
  W(\omega)=
  \psi_\mathrm{L}(s_0,\omega)\psi_\mathrm{R}'(s_0,\omega)
  -\psi_\mathrm{L}'(s_0,\omega)\psi_\mathrm{R}(s_0,\omega)\;,
\eeql{W}
which has the standard property of being independent of the choice of matching point~$s_0$.\footnote{For an $s$-contour consisting of two straight lines, the point where they join is the obvious choice.} At the zeroes of $W(\omega)$, which can be found with a standard complex-root solver,\footnote{Minimization of $|W(\omega)|^2$ should be a feasible elementary strategy: not only does $|W|^2$ clearly attain a minimum at zeroes of~$W$, but the standard maximum/minimum modulus theorem of complex analysis (readily verified using a Taylor expansion of~$W$) guarantees that these are the \emph{only} minima.} $\psi_\mathrm{L}$ and $\psi_\mathrm{R}$ are linearly dependent, which means that either function satisfies \emph{both} boundary conditions, i.e., $\omega$ is an eigenvalue. An advantage of (B) is that the numerical determinations of $\psi_\mathrm{L}$ and $\psi_\mathrm{R}$ may be optimized separately. The Wronskian method is also particularly suitable when only a few resonances are needed, and their approximate location is known. This is the case here, for one is only interested in the few ``physical'' resonances corresponding to the finite number of metastable well states.\footnote{There is, however, a theoretical interest in the fate of these resonances as they ``fall out of the well" when $\alpha$ is lowered. Note that the resonance frequencies, being complex zeroes of the analytical function $W(\omega)$, should in all cases depend continuously on the system parameters according to the argument principle.}

Let us finally point out that formulating the problem on a Stokes contour not only makes $\psi_\mathrm{L,R}$ well-defined, but also renders their numerical calculation very accurate and stable. Namely, choosing initial points $s_1$ with $|s_1|\gg1$, any error in the initial conditions\footnote{From (\ref{psiR}) and above, $\psi_\mathrm{L}'(s_1)=(-s_1)^{3/2}\psi_\mathrm{L}(s_1)$ and $\psi_\mathrm{R}'(s_1)=is_1^{3/2}\psi_\mathrm{R}(s_1)$ are seen to be suitable starting values.} amounts to an admixture of the undesired dominant solution. The latter is exponentially suppressed when integrating ``inward" from the asymptotic point $s_1$ to the matching point~$s_0$, while the desired subdominant solution is amplified by a similar factor. The appendix illustrates this in an elementary setting.

\subsection{Analytics}
\label{analytics}

For the potential $U(s)=\alpha s-s^3$, the well minimum is at $s_\mathrm{w}=-\sqrt{\alpha/3}$. From $U''(s_\mathrm{w})=\half\op^2$ one finds $\op=2(3\alpha)^{1/4}$ for the plasma frequency. By inversion symmetry, the barrier height is $\Delta U=-2U(s_\mathrm{w})=\frac{4}{9}\sqrt{3\alpha^3}$. For the number of states in the well, one now has the estimate $N_\mathrm{s}=\Delta U/\op=\frac{2}{9}\sqrt[4]{3\alpha^5}$; if $N_\mathrm{s}\gg1$, one is within the semiclassical regime, at least for the lowest few states $n=0,1,2,\ldots$. In this case, one can refer to the literature for a prediction for the resonance eigenvalues:
\beq
  \omega_n\approx U(s_\mathrm{w})+(n{+}\half)\op+\omega_n^\mathrm{anh}
           -{\textstyle\frac{i}{2}}\Gamma_n(\op,\Delta U)\;.
\eeql{semi}
Here, $\omega_n^\mathrm{anh}$ is a small (in the semiclassical limit) correction due to the anharmonicity of the well, the lowest few of which are known explicitly: $\omega_0^\mathrm{anh}=-11\op^{-4}$, $\omega_1^\mathrm{anh}=-71\op^{-4}$, and $\omega_2^\mathrm{anh}=-191\op^{-4}$~\cite{ajb}. The decay rate is~\cite{CL,rate}
\beq
  \Gamma_n=\op\frac{{(432N_\mathrm{s})}^{n+1/2}}{\sqrt{2\pi}n!}
           \,e^{-36N_\mathrm{s}/5}\;;
\eeq
it occurs in (\ref{semi}) with a prefactor $\half$ due to the conversion from probabilities to amplitudes. Combining the above, the prediction (\ref{semi}) is seen to be completely explicit in terms of~$\alpha$; it should be valid whenever $n\ll N_\mathrm{s}$.

The asymptotics around (\ref{psiR}) have an interesting consequence: $\psi_\mathrm{L}(s)\sim e^{-\frac{2}{5}(-s)^{5/2}}$ can be extended to $\frac{2}{5}\pi<\arg s<\frac{8}{5}\pi$, and $\psi_\mathrm{R}(s)\sim e^{i\frac{2}{5}s^{5/2}}$ to $-\frac{2}{5}\pi<\arg s<\frac{4}{5}\pi$. Hence, there is a region of overlap, and for an eigenfunction $\psi_n(s)$, we conclude that the ``left" and ``right" asymptotic forms should occur with the \emph{same} prefactor. Note that, for $\frac{2}{5}\pi<\arg s<\frac{4}{5}\pi$, we have compared two dominant asymptotic solutions, which can be done only up to a subdominant solution. Therefore, we first have to ascertain by other means that a solution satisfying both boundary conditions actually exists, and asymptotics alone cannot yield the eigenvalues~$\omega_n$. In particular, if a solution, say $\psi_\mathrm{L}(s{\To}{-}\infty,\omega)\sim1\cdot e^{-\frac{2}{5}(-s)^{5/2}}$, is extended to $s>0$ for $\omega\in\mathbb{R}$, i.e., $\psi_\mathrm{L}(s{\To}{+}\infty,\omega)\sim ae^{i\frac{2}{5}s^{5/2}}+a^*e^{-i\frac{2}{5}s^{5/2}}$, then it seems that the preceding argument \emph{cannot} determine~$a$, because the last asymptotic expansion is valid only on $\lvert\arg s\rvert<\frac{2}{5}\pi$ due to the Stokes phenomenon. Of course, these claims should be verified numerically.

For an alternative approach to (\ref{psiR}), consider the Riccati formulation. If $\psi''=[U-\omega]\psi$, then
\beq
  f'+f^2=U-\omega
\eeql{Ric}
for the \emph{logarithmic derivative} $f=\psi'\!/\psi$. Thus, a second-order linear equation has been converted to a first-order nonlinear one. This has several advantages. First, the linearity of~(\ref{scale}), while a blessing in many contexts, also means that solutions are arbitrary up to an overall constant; this ambiguity cancels in the definition of~$f$. For instance, the Riccati version of the matching criterion reads simply $f_\mathrm{L}=f_\mathrm{R}$. Also, the WKB approximation is obvious by neglecting $f'$ in~(\ref{Ric}). More importantly, the exponentially large or small tails typically present in $\psi$ cancel in~$f$, so that one deals with at most algebraically large quantities. A severe disadvantage of (\ref{Ric}) usually is that it converts the nodes of $\psi$ into simple poles of~$f$. However, while nodes are ubiquitous for the real $\psi$ one usually encounters for real $s$ and~$\omega$, presently these nodes are readily circumvented in the $s$-plane.

The above leads to slightly different recommendations for the integration contour $\mathcal{C}$ in the Schr\"odinger and Riccati formulations. For (\ref{scale}), one could either leave the real $s$-axis at $s=0$ (upon which yet another reparametrization $s=-re^{i\pi/5}$ reduces the inward problem for $\psi_\mathrm{R}$ to standard form) to keep the parametrization as simple as possible, or stay on this axis at least until the barrier point $s=|s_\mathrm{w}|$, so that the (complex) resonance wave function is as amenable as possible to a conventional interpretation. For (\ref{Ric}), on the other hand, one should leave the real $s$-axis at an $s<0$ to the left of the classically allowed region, where $\psi$ will have (near-)nodes on the real axis.

Some familiar properties of quantum mechanics can be generalized to the present setting. Consider wave functions $\psi$ and $\chi$, each satisfying both boundary conditions on~$\mathcal{C}$, and introduce their bilinear product $(\psi,\chi)\equiv\int_\mathcal{C}ds\,\psi(s)\chi(s)$. One now derives $(\psi,H\chi)=(\chi,H\psi)$, through the usual integration by parts. In particular, for eigenfunctions this implies $(\omega_m{-}\omega_n)(\psi_m,\psi_n)=0$, or
\beq
  (\psi_m,\psi_n)=0\;,\qquad m\neq n\;,
\eeq
where we also used that the 1D Schr\"odinger equation has no degeneracies, so $m\neq n$ if $\omega_m\neq\omega_n$. The derivation of this ``orthogonality" relied on the symmetry of~$H$; for the latter, it is essential that the bilinear product does not involve complex conjugation, since $H$ is not real on~$\mathcal{C}$. However, this has the consequence that the diagonal entries $(\psi,\psi)$ are not positive definite, not even necessarily real or nonzero~\cite{JB}, which is why we refrained from normalizing ``$(\psi_m,\psi_n)=\delta_{mn}$". Also, there is no suggestion of completeness.

For future reference we introduce the Green function $G(s,u;\omega)=G(u,s;\omega)$ through
\beq
  G(s,u;\omega)=
  \frac{\psi_\mathrm{L}(s,\omega)\psi_\mathrm{R}(u,\omega)\theta(u{-}s)+
        \psi_\mathrm{L}(u,\omega)\psi_\mathrm{R}(s,\omega)\theta(s{-}u)}
       {W(\omega)}\quad(s,u\in\mathbb{R})\;,
\eeql{G}
solving $[H_\mathrm{S}(s)-\omega]G(s,u;\omega)=-\delta(s{-}u)$. Since $W(\omega)$ has zeroes at the resonances~$\omega_n$, these correspond to poles of~$G$, and the eigenfunctions $\psi_n$ are the associated residues~\cite{RMP}. A major feature of this work is the generalization to complex~$s$, so that the definition of $G$ potentially leads to functions of two complex variables. This is emphatically avoided, and below we will at most consider $G(s,u;\omega)$ with $s$ and $u$ on the same fixed contour~$\mathcal{C}$ (having a real parametrization), in which case the generalization of (\ref{G}) is still obvious.

For an application of the above, let us consider the response to changes in~$\alpha$. This provides an opportunity to acquaint oneself with the perturbation theory required in the interacting case, but in a scalar (as opposed to spinor) setting. Thus, consider the $\alpha$-derivative of (\ref{scale}),
\beq
  -\xi''+(\alpha s-s^3-\omega)\xi+(s-\omega')\psi=0\;,
\eeql{xi-eqn}
where $\xi(s,\alpha)\equiv\pt_\alpha\psi(s,\alpha)$ and $\omega'(\alpha)=d\omega(\alpha)/d\alpha$ without risk of confusion, and where $\xi$, $\psi$, and $\omega$ still have an implicit dependence on the discrete mode index~$n$. Clearly, (\ref{xi-eqn}) defines $\xi$ at most up to a multiple of~$\psi$, reflecting the freedom to choose an $\alpha$-dependent normalization constant in~$\psi(\alpha)$. Rewriting (\ref{xi-eqn}) as $[(\xi/\psi)'\psi^2]'=(s-\omega')\psi^2$, one sees that the integral of the lhs must vanish over the integration contour~$\mathcal{C}$, since $\psi$ obeys nodal conditions at both ends for any~$\alpha$, and therefore so does~$\xi$. Hence, the same must hold for the integral of the rhs, i.e.,
\beq
  \frac{d\omega}{d\alpha}=
  \frac{\int_\mathcal{C}\!ds\,s\psi^2(s)}{\int_\mathcal{C}\!ds\,\psi^2(s)}
  =\frac{(\psi,s\psi)}{(\psi,\psi)}\equiv\langle s\rangle\;.
\eeql{alpha-pert}
This relation gives the $\alpha$-response in a familiar form, as the normalized matrix element of the perturbing operator~$s$. However, the integrals are evaluated on the contour $\mathcal{C}$ joining two Stokes lines, ensuring rapid convergence, and importantly involve $\psi^2$ not~$|\psi|^2$. Note also that (\ref{alpha-pert}) simultaneously gives the real and imaginary parts of the shift.

\subsection{Numerical results}

So far, the integration of $\psi_\mathrm{L,R}(s,\omega)$ has been brought under control, and the exploration of the resulting $W(\omega)$ has commenced. Further results are expected shortly.

\section{Interacting problem}
\label{int}

The transformation leading to (\ref{scale}) also applies to the general $H$ in~(\ref{H}), and scaled parameters $(\tilde{\epsilon},\tilde{\Delta},\tilde{\lambda})$ are readily written down in analogy to $\tilde{\alpha}$,~$\tilde{\omega}$. However, after this, no further exact simplification is apparent for~$H$, so that the interacting problem is considerably richer than the uncoupled one. However, there is an alternative form
\beq
  \hat{H}=H_\mathrm{S}-\half\Omega\sigma_z+(\lambda_z\sigma_z{+}\lambda_x\sigma_x)s\;,
\eeql{H1}
in which the free qubit Hamiltonian, not the interaction, is written in diagonal form. Specifically, defining $\cot\eta=\epsilon/\Delta$, $0\le\eta\le\pi$, one has $\hat{H}=e^{i\eta\sigma_y/2}He^{-i\eta\sigma_y/2}$, provided that $\lambda_z=\lambda\cos\eta$ and $\lambda_x=-\lambda\sin\eta$. Thus, the spectra of $H$ and $\hat{H}$ are identical, and their eigenfunctions trivially related.

\subsection{Analytics}
\label{int-analytics}

The above, while seemingly trivial, has an intriguing consequence for the analytical study of~(\ref{SE}), which now becomes a well-defined problem in resonance perturbation theory. Routinely following the standard prescription ``$H=H_\text{qb}+H_\text{readout}+H_\text{int}$'', one is led to expand in $H_\text{int}$, i.e., in~$\lambda$. However, (\ref{SE}) is also ``solvable" (in terms of the resonances of~$H_\mathrm{S}$) for finite $\lambda$ and $\Delta=0$, in which case the spinor components decouple. By (\ref{H1}), we see that a $\Delta$\nobreakdash-expansion assumes $\lambda_x\To0$ at arbitrary~$\lambda_z$, while the $\lambda$-expansion assumes \emph{both} $\lambda_x,\lambda_z\To0$. Conceptually, $\hat{H}$ is appealing: it shows that the QND component $\lambda_z$ of the coupling can be readily handled, while the non-QND (henceforth QD) component $\lambda_x$ is nontrivial.

In contrast, the $\lambda$-expansion, which mixes both components, places an unnecessary restriction on the QND component, and obscures the fact that a QD perturbation (either $\lambda_x$ or~$\Delta$) affects the resonance frequencies only in second order.\footnote{Note that the effect of $\lambda_x$ or~$\Delta$ must be even, since their sign is a matter of convention, or draw a picture on the Bloch sphere.} Still, we give for reference
\beq
  \left.\frac{\pt\omega(\epsilon,\Delta,\lambda)}{\pt\lambda}\right|_{\lambda=0}
  =\cos\eta\left.\frac{\pt\hat{\omega}(\Omega,\lambda_z,\lambda_x{=}0)}
                      {\pt\lambda_z}\right|_{\lambda_z=0}\;,
\eeql{l-exp}
where we used $\pt\hat{\omega}(\Omega,\lambda_z{=}0,\lambda_x)/ \pt\lambda_x|_{\lambda_x=0}=0$. For the qubit in its ground state, one has $\cos\eta=\langle\sigma_z\rangle$ and $\pt\omega(\Omega,\lambda_z,\lambda_x{=}0)/\pt\lambda_z|_{\lambda_z=0}=\omega_\mathrm{S}'$, with $\omega_\mathrm{S}'$ as in (\ref{alpha-pert}), where quantities pertaining to the uncoupled SQUID $H_\mathrm{S}$ will be henceforth denoted with an ``S" subscript. For an excited qubit, both these relations have the opposite sign. Hence, in either case, one obtains $\delta\omega=\lambda\omega_\mathrm{S}'\langle\sigma_z\rangle$ for the first-order frequency shift (where on the rhs only $\omega_\mathrm{S}'$ has an imaginary part): self-averaging (\emph{Hamiltonian-dominated}~\cite{MSS}) measurement is a direct consequence of the weak-coupling assumption. The SQUID's working point $\alpha$ strongly affects the measurement sensitivity through~$\omega_\mathrm{S}'$.

Conversely, let us consider the first-order effect of the qubit splitting $\Omega$ in the presence of a finite coupling~$\lambda$. For only QND coupling, the qubit spin always has a definite value, which we denote by $s_z$ for $H$ and $\hat{s}_z$ for $\hat{H}$, so that $\hat{s}_z=s_z\sgn\epsilon=s_z\sgn\lambda_z$. One now has
\beq\begin{split}
  \omega(\epsilon,\Delta,\lambda)|_{\Omega\To0}&\approx\omega(0,0,\lambda)
  +\epsilon\frac{\pt\omega(0,0,\lambda)}{\pt\epsilon}
  +\Delta\frac{\pt\omega(0,0,\lambda)}{\pt\Delta}\\
  &=\omega_\mathrm{S}(\alpha{+}\lambda s_z)-\frac{\epsilon s_z}{2}\;,
\end{split}\eeql{det-dom}
where on the first line, the second term is trivial since it is evaluated in the absence of tunneling, and the third term has been argued above to cancel. One can interpret (\ref{det-dom}) by saying that strong coupling to the SQUID has collapsed the qubit in the pointer basis ($\sigma_z$~eigenbasis) and suppressed tunneling---\emph{detector-dominated} measurement~\cite{MSS}.

For a systematic approach, one should relate and work out the $\Delta$- and $\lambda_x$-expansions analogously to~(\ref{l-exp}):
\beq\begin{split}
  {\pt_\eta^2\omega(\Omega\cos\eta,\Omega\sin\eta,\lambda)|}_{\eta=0,\pi}&=
  {\pt_\eta^2\hat{\omega}(\Omega,\lambda\cos\eta,-\lambda\sin\eta)|}_{\eta=0,\pi}\\
  &=\frac{\epsilon s_z}{2}
    +\epsilon^2\left.\frac{\pt^2\omega}{\pt\Delta^2}\right|_{\Delta=0}\\
  &=-\lambda_z\hat{s}_z\omega_\mathrm{S}'(\alpha{+}\lambda_z\hat{s}_z)
    +\lambda_z^2\left.\frac{\pt^2\hat{\omega}}{\pt\lambda_x^2}\right|_{\lambda_x=0}
  \;.
\end{split}\eeql{exp-rel}
That is, in general the two schemes yield equivalent information, but the $\Delta$-expansion becomes trivial for $\lambda=0$ (no coupling in any order), while the $\lambda_x$-expansion becomes trivial for $\Omega=0$ (QND in any order), with of course $\pt_\eta^2\omega=0$ in either case since the problem is $\eta$-independent unless both $\lambda$ and $\Omega$ are nonzero. Note also that, from its prefactor $\epsilon^2$ in~(\ref{exp-rel}), one cannot conclude that $\pt_\Delta^2\omega|_{\Delta=0}$ diverges for $\epsilon\To0$, because in this limit the two terms on the last line of (\ref{exp-rel}) cancel each other. Rather, $\pt_\Delta^2\omega|_{\Delta=\epsilon=0}$ should be calculated directly from (\ref{D-exp}) below, and cannot be trivially related to other expansion coefficients. Analogous remarks apply to $\pt_{\lambda_x}^2\hat{\omega}|_{\lambda_x=\lambda_z=0}$.

Next, the actual expansion should be carried out. This is sketched for the $\Delta$-expansion of the resonance with $s_z=1$; other cases are analogous. In $\Oc(\Delta^0)$, $\psi_\ua^{(0)}(\alpha,\epsilon,\lambda)=\psi_\mathrm{S}(\alpha{+}\lambda)$ and $\psi_\da^{(0)}=0$, while $\omega^{(0)}(\alpha,\epsilon,\lambda)= \omega_\mathrm{S}(\alpha{+}\lambda)-\half\epsilon$. In $\Oc(\Delta)$, the coupling of the spinor components means that $\psi_\ua^{(0)}$ is a source term in the equation for $\psi_\da^{(1)}$, which is solved as
\beq
  \psi_\da^{(1)}(s;\alpha,\epsilon,\lambda)=-\half\int_\mathcal{C}\!du\,
  G\bm{(}s,u;\omega_\mathrm{S}(\alpha{+}\lambda){-}\epsilon;\alpha{-}\lambda\bm{)}
  \,\psi_\mathrm{S}(u,\alpha{+}\lambda)\;,
\eeq
where the Green function (\ref{G}) of the free SQUID is simply denoted as $G$ since there is no risk of confusion. Finally, $\psi_\da^{(1)}$ in its turn figures as a source term in the equation for $\psi_\ua^{(2)}$, from which the frequency shift can be determined in much the same way as in~(\ref{alpha-pert}). The final results are
\begin{align}
  \left.\frac{\pt^2\omega}{\pt\Delta^2}\right|_{\Delta=0}&=\frac{
  \int_\mathcal{C}\!ds\,du\,\psi_\mathrm{S}(s,\alpha{+}\lambda s_z)\,
  G\bm{(}s,u;\omega_\mathrm{S}(\alpha{+}\lambda s_z){-}\epsilon s_z;\alpha{-}\lambda s_z\bm{)}
  \,\psi_\mathrm{S}(u,\alpha{+}\lambda s_z)}
  {2\int_\mathcal{C}\!ds\,\psi_\mathrm{S}^2(s,\alpha{+}\lambda s_z)}\;,\label{D-exp}\\
  \left.\frac{\pt^2\hat{\omega}}{\pt\lambda_x^2}\right|_{\lambda_x=0}&=2\,\frac{
  \int_\mathcal{C}\!ds\,du\,s\psi_\mathrm{S}(s,\alpha{+}\lambda_z\hat{s}_z)\,
  G\bm{(}s,u;\omega_\mathrm{S}(\alpha{+}\lambda_z\hat{s}_z){-}\Omega\hat{s}_z;\alpha{-}\lambda_z\hat{s}_z\bm{)}
  \,u\psi_\mathrm{S}(u,\alpha{+}\lambda_z\hat{s}_z)}
  {\int_\mathcal{C}\!ds\,\psi_\mathrm{S}^2(s,\alpha{+}\lambda_z\hat{s}_z)}\;.\label{Lx-exp}
\end{align}
There seems to be an interesting resonant enhancement if $\omega_\mathrm{S}^{(j)}(\alpha{+}\lambda s_z){-}\epsilon s_z\approx\omega_\mathrm{S}^{(k)}(\alpha{-}\lambda s_z)$, where the SQUID initially is in mode~$j$, while $k$~denotes another mode. Namely, in this case $G$ in (\ref{D-exp}) is evaluated near a pole. However, caution is needed, for this ``resonance'' condition could be met even if $\lambda=0$, where one knows that (\ref{D-exp}) in fact is trivial.

Note that, \emph{if} the free SQUID's eigenfunctions were complete, the Green function (or resolvent) would have a representation of the form $G(s,u;\omega)\sim\sum_n\psi_{\mathrm{S},n}(s)\psi_{\mathrm{S},n}(u)/(\omega{-}\omega_n)$. In such a case, (\ref{D-exp}) and (\ref{Lx-exp}) would reduce to a more familiar ratio of squared matrix elements and ``energy denominators", summed over the mode index~\cite{RMP}.

It is instructive to verify the predicted equivalence~(\ref{exp-rel}) directly from these integral expressions. For that purpose, we write (\ref{scale}) as
\beq
  -\psi_\mathrm{S}(s,\alpha)''+[\alpha_1s-s^3-\omega_1]\psi_\mathrm{S}(s,\alpha)=
  [\omega_\mathrm{S}(\alpha)-\omega_1+(\alpha_1{-}\alpha)s]\psi_\mathrm{S}(s,\alpha)\;,
\eeq
for \emph{arbitrary} $\alpha_1,\omega_1\in\mathbb{R}$. One can invert the differential operator on the lhs, yielding
\beq
  \psi_\mathrm{S}(s,\alpha)=\int_\mathcal{C}\!du\,G(s,u;\omega_1;\alpha_1)
  [\omega_1-\omega_\mathrm{S}(\alpha)+(\alpha{-}\alpha_1)u]\psi_\mathrm{S}(u,\alpha)\;.
\eeql{G-rel}
Applying (\ref{G-rel}) twice to (\ref{Lx-exp}), first to the $u$-integral and then to the $s$-integral, one verifies that (\ref{D-exp}) and (\ref{Lx-exp}) are indeed related by~(\ref{exp-rel}) (in particular, also when one of the two expressions becomes trivial).~$\square$

The above represents substantial formal progress, but further analysis is needed to elucidate the physics contained in (\ref{D-exp}) and (\ref{Lx-exp}). Presumably, one should try to evaluate these, in particular their imaginary parts, in the semiclassical limit $n\ll N_\mathrm{s}$, and ideally make a comparison with nonperturbative numerical results for the interacting system.

\subsection{Degenerate perturbation theory of the linear-detector regime}
\label{int-deg}

The above allowed for an arbitrary angle $\eta$ between the energy and pointer bases only in two disjoint cases: infinitesimal $\lambda$ as in (\ref{l-exp}), or infinitesimal $\Omega$ as in (\ref{det-dom}). One is led to suspect that a unified treatment is possible if \emph{both} $\lambda$ and $\Omega$ are small, but this regime cannot be directly extracted from the above. Namely, in the leading order $\lambda=\Omega=0$, the problem reduces to the free SQUID, and hence is degenerate in the spin variable. One also notices that the central formulae (\ref{D-exp}) and (\ref{Lx-exp}) become singular in this limit, for the Green functions $G$ are needed infinitesimally close to their resonance frequency, where they diverge.

Therefore, here we give a separate calculation, expanding in $\epsilon$, $\Delta$, and $\lambda$ simultaneously. Physically, this (presumably) means that both the qubit splitting and the typical interaction energy are very small compared to the SQUID plasma frequency. In its turn, this means that only one of its (noninteracting) resonances is needed to describe the SQUID (typically the longest-lived ``ground-state" one), i.e., that the SQUID follows the qubit adiabatically---all in all, adiabatic elimination is just a form of perturbation expansion. In the first order, which will be considered here, this moreover agrees with the \emph{linear-detector regime}, when the qubit perturbs the detector only weakly, without necessarily assuming the reverse.

Typically, degenerate perturbation theory starts with considering the matrix elements of the perturbing operator in the degenerate subspace. Here, some care is needed since there is no concept of a basis; however, one can proceed in analogy with (\ref{alpha-pert}), and the product $(\bm{\cdot},\bm{\cdot})$ will emerge naturally. Thus, consider the leading correction to (\ref{scale}) analogous to (\ref{xi-eqn}),
\beq
  [H_\mathrm{S}-\omega_\mathrm{S}(\alpha)]\ket{\Xi}+[-\delta\omega
  -\half(\epsilon\sigma_z{+}\Delta\sigma_x)+\lambda\sigma_zs]\ket{\Psi}=0\;,
\eeql{Xi}
where $\delta\omega$ is the leading correction to the resonance frequency~$\omega_\mathrm{S}(\alpha)$, $\ket{\Xi}=(\xi_\ua,\xi_\da)^\top$ is the corresponding correction to the lowest-order wavefunction $\ket{\Psi}=\bigl(\begin{smallmatrix} p \\ q \end{smallmatrix}\bigr)\psi_\mathrm{S}$, and where $p,q$ are as yet undetermined. As before, $H_\mathrm{S}-\omega_\mathrm{S}(\alpha)$ is not invertible, so a solution $\ket{\Xi}$ satisfying both boundary conditions only exists in a special case, namely if
\beq
  \begin{pmatrix} -\half\epsilon{+}\lambda\omega_\mathrm{S}' & -\half\Delta \\
                  -\half\Delta & \half\epsilon{-}\lambda\omega_\mathrm{S}' \end{pmatrix}
  \begin{pmatrix} p \\ q \end{pmatrix} = \delta\omega \begin{pmatrix} p \\ q \end{pmatrix}.
\eeql{key}
This is a key result: the resonances $\delta\omega$ follow from a matrix equation, involving both the qubit parameters $\epsilon,\Delta$ and the \emph{complex} linearized detector response~$\omega_\mathrm{S}'$, as $\delta\omega=\pm\half\sqrt{(\epsilon{-}2\lambda\omega_\mathrm{S}')^2+\Delta^2}$. Depending on the relative magnitudes of $\Omega$ and $\lambda|\omega_\mathrm{S}'|$, the Hamiltonian- and detector-dominated regimes are now trivially extracted, as will be sketched below for the purpose of interpretation.

First, however, let us pursue the determination of~$\ket{\Xi}$. Assuming that $p$, $q$, and $\delta\omega$ are related as in~(\ref{key}), one can rewrite (\ref{Xi}) as
\beq
   [H_\mathrm{S}-\omega_\mathrm{S}(\alpha)]\ket{\Xi}
   +\lambda(s-\langle s\rangle)\sigma_z\ket{\Psi}=0\;,
\eeql{Xi2}
and comparison with (\ref{xi-eqn}) yields the solution $\ket{\Xi}=\lambda\sigma_z\pt_\alpha\ket{\Psi}$. That is, to this order, the corrected wavefunction can be written as
\beq
  \ket{\Psi_\lambda}\approx\ket{\Psi}+\ket{\Xi}\approx
  \begin{pmatrix} p\psi_\mathrm{S}(\alpha{+}\lambda) \\ q\psi_\mathrm{S}(\alpha{-}\lambda)
  \end{pmatrix}.
\eeql{Xi-res}
In agreement with the above adiabatic picture, (\ref{Xi-res})~gives the SQUID wavefunction as the instantaneous \emph{eigen}function corresponding to the qubit flux state. In particular, and unsurprisingly, the system's wavefunction is essentially always entangled, with the degree of entanglement characterized by the coupling~$\lambda$. The exception are eigenstates with $p\To0$ or $q\To0$, that is, QND or detector-dominated measurement of a classical flux state.

Let us discuss the two limiting cases. In the detector-dominated regime $\Omega\ll\lambda|\omega_\mathrm{S}'|$, one finds $\delta\omega_0\approx\lambda\omega_\mathrm{S}'-\half\epsilon +\Delta^2\!/8\lambda\omega_\mathrm{S}'$ and ${(p,q)}_0\propto(1,-\Delta/4\lambda\omega_\mathrm{S}' -\Delta\epsilon/8(\lambda\omega_\mathrm{S}')^2)$, while $\delta\omega_1=-\delta\omega_0$ and ${(p,q)}_1\propto(\Delta/4\lambda\omega_\mathrm{S}' +\Delta\epsilon/8(\lambda\omega_\mathrm{S}')^2,1)$. That is, the qubit has ``collapsed" in the pointer basis; the frequencies are mainly determined by the detector response and the bias~$\epsilon$, while tunneling $\sim\Delta^2$ is strongly suppressed with a factor $(\lambda\omega_\mathrm{S}')^{-1}$. Conversely, in the Hamiltonian-dominated regime $\Omega\gg\lambda|\omega_\mathrm{S}'|$, one has $\delta\omega\approx\pm[\half\Omega-\lambda\omega_\mathrm{S}'\epsilon/\Omega +(\lambda\omega_\mathrm{S}')^2\Delta^2\!/\Omega^3$]. The first two terms are $\pm\half\Omega+\lambda\omega_\mathrm{S}'\langle\sigma_z\rangle$, familiar from below~(\ref{l-exp}). Rewriting the last term as $\delta\omega=\cdots\pm(\lambda_x\omega_\mathrm{S}')^2/\Omega$, we tentatively interpret as follows: for $\omega_\mathrm{S}'$ mostly real, $\lambda_x$ causes SQUID-assisted tunneling, increasing the effective tunnel splitting. For $\omega_\mathrm{S}'$ mostly imaginary, however, the measurement of $\sigma_z$ may be said to cause a Zeno effect, reducing the effective tunnel splitting. The corresponding wavefunctions are readily written down when needed.

As clearly as (\ref{key}) shows the transition from $H$-dominated to detector-dominated measurement, it does not address all conceptual questions. Namely, it always describes weak measurement; on the opposite side, one has single-shot readout. ``Single-shot" means that for a suitable waiting time, decay for one state is almost complete, while for the other state it is negligible, i.e., $\lvert\im\omega_0\rvert\ll\lvert\im\omega_1\rvert$. In general, one presumably should introduce a discriminating power based on the ratio of these decay rates. In this framework, one can study issues such as whether single-shot measurement can be $H$-dominated, and the existence of a dimensionless quantum efficiency $0\le\eta\le1$ like for continuous measurement.

\subsection{Numerics}
\label{int-num}

Since the uncoupled Schr\"odinger equation for $H_\mathrm{S}$ already is implemented numerically as two coupled first-order equations, the integration of $\ket{\Psi}$ should only involve a trivial generalization. Also, the cubic potential term continues to dominate asymptotically in the presence of a qubit, so the boundary conditions and choice of integration contour $\mathcal{C}$ carry over unchanged. The matching, however, needs more care. Based on the trivial cases $\lambda=0$ or $\Delta=0$, one now expects \emph{two} linearly independent solutions for each of $\ket{\Psi_\mathrm{L}}$ and $\ket{\Psi_\mathrm{R}}$ satisfying the corresponding boundary conditions. During the numerical integration, we have no \emph{a priori} knowledge which solution is the correct one, so we have to integrate both, and determine a \emph{fundamental system} $\{\ket{\Psi_\mathrm{L}^{(1)}},\ket{\Psi_\mathrm{L}^{(2)}}\}$ of independent spinors, obeying $\ket{\Psi_\mathrm{L}^{(j)}(s)}\To0$ for $s\To-\infty$ ($j=1,2$), and analogously for $\mathrm{L}\leftrightarrow\mathrm{R}$. Subsequently, the question becomes for which $\omega$ is there a superposition of the $\ket{\Psi_\mathrm{L}^{(j)}}$ matching onto a superposition of the $\ket{\Psi_\mathrm{R}^{(j)}}$, of course for both spinor components and for the first derivatives as well as the function values. In other words, one has to find the zeroes of the \emph{Wronski determinant}
\beq
  W(\omega)=\left|\begin{array}{cccc}
   \psi_{\ua\mathrm{L}}^{(1)}(s_0,\omega) &
   \psi_{\ua\mathrm{L}}^{(2)}(s_0,\omega) &
   \psi_{\ua\mathrm{R}}^{(1)}(s_0,\omega) &
   \psi_{\ua\mathrm{R}}^{(2)}(s_0,\omega) \\[1mm]
   \psi_{\da\mathrm{L}}^{(1)}(s_0,\omega) &
   \psi_{\da\mathrm{L}}^{(2)}(s_0,\omega) &
   \psi_{\da\mathrm{R}}^{(1)}(s_0,\omega) &
   \psi_{\da\mathrm{R}}^{(2)}(s_0,\omega) \\[1mm]
   \psi_{\ua\mathrm{L}}^{(1)\prime}(s_0,\omega) &
   \psi_{\ua\mathrm{L}}^{(2)\prime}(s_0,\omega) &
   \psi_{\ua\mathrm{R}}^{(1)\prime}(s_0,\omega) &
   \psi_{\ua\mathrm{R}}^{(2)\prime}(s_0,\omega) \\[1mm]
   \psi_{\da\mathrm{L}}^{(1)\prime}(s_0,\omega) &
   \psi_{\da\mathrm{L}}^{(2)\prime}(s_0,\omega) &
   \psi_{\da\mathrm{R}}^{(1)\prime}(s_0,\omega) &
   \psi_{\da\mathrm{R}}^{(2)\prime}(s_0,\omega)
  \end{array}\right|.
\eeql{W-det}
Ignoring the spin index, the form~(\ref{W}) for the uncoupled Wronskian is seen to be precisely the $\bigl(\begin{smallmatrix} (1,1) & (1,3) \\ (3,1) & (3,3) \end{smallmatrix}\bigr)$ minor determinant of~(\ref{W-det}).

\section*{Acknowledgment}

I thank M.H.S. Amin, A.J. Berkley, and A.Yu.\ Smirnov for discussions, and K.~Young for a long collaboration which introduced me to quasinormal modes. S.~Cronin contributed to the numerical analysis in an early stage, finding evidence for the existence of resonance solutions.

\appendix

\section{Toy model}

Consider a particle with mass $m=\half$ on the positive $x$-axis,
\beq
  -\psi''(x)+U(x)\psi(x)=\omega\psi(x)\;.
\eeql{toy}
The origin $x=0$ is supposed to be a hard wall, yielding the boundary condition $\psi(0)=0$. For $x>0$, the potential is a finite well, $U(x)=V\theta(a{-}x)$, with $\theta$ the unit step function and $V<0<a$. Depending on $V,a$, the problem has a finite number (possibly zero) of bound states with $V<\omega<0$. These are to be determined by the Wronskian method of~(\ref{W}), with $x=a$ being the obvious choice for the matching point. On $0\le x\le a$, the ``left" function $\psi_\mathrm{L}(x,\omega)=\sin(\sqrt{\omega{-}V}x)/\sqrt{\omega{-}V}$ is readily determined from the nodal condition. The same is true for the ``right" function $\psi_\mathrm{R}(x,\omega)=e^{-\sqrt{-\omega}x}$ on $x>a$, but only because the closed-form solution enables us to deal with the singularity at $x=+\infty$; instead, one can examine how ``inward" integration works in this example. \emph{First}, pick some trial initial values, say $\psi_\mathrm{R}(x_0,\omega)=-\psi_\mathrm{R}'(x_0,\omega)=1$; in general, $\{\psi_\mathrm{R}(x_0,\omega),\,\psi_\mathrm{R}'(x_0,\omega)\}$ could be left undetermined. \emph{Second}, solve the Schr\"odinger equation (\ref{toy}) on $x>a$ with these trial initial values, e.g.\ normalizing the result such that $\psi_\mathrm{R}(a,\omega)=1$. [In fact, as usual, the logarithmic derivative $\psi_\mathrm{R}'(a,\omega)/\psi_\mathrm{R}(a,\omega)$ is all that matters for Wronskian matching.] For the toy model, this can be done analytically, while in general it would be a numerical task. \emph{Third}, consider $\psi_\mathrm{R}$ as $x_0$ becomes large, while the trial initial values $\{\psi_\mathrm{R}(x_0,\omega),\,\psi_\mathrm{R}'(x_0,\omega)\}$ are kept fixed. One observes that the error committed by choosing a finite $x_0$ decreases exponentially.

In slightly more technical language, the above shows that, even though $x=\infty$ is an irregular singular point of the Schr\"odinger equation for a free particle, one can still effectively implement the boundary condition $\psi_\mathrm{R}(x{=}{+}\infty,\omega{<}0)=0$ numerically.

To complete this exercise, one could now perform the matching and calculate approximate eigenvalues $\{\tilde{\omega}_j(x_0)\}$, studying their convergence to the true spectrum $\{\omega_j\}$ as $x_0\To\infty$. Here, (\ref{toy}) is instead pursued analytically, as a toy model for resonance states of the Schr\"odinger equation. For $V<0$ as above, this is illuminating, since the bound states now feature in \emph{both} types of spectrum of the system: they are the discrete part of the conventional real hermitian-operator spectrum, and the undamped members of the resonance spectrum. For $(\re)\,\omega>0$, the continuous conventional spectrum and the damped discrete resonances offer complementary descriptions, and correspond to different definitions of an eigenvalue. Moreover, the model can be considered for $V>0$ as well, in which case there are also resonances. This clarifies that the resonances are caused by wave scattering rather than by classical confinement: in quantum mechanics, confinement is not absolute at energies below the pertinent barrier energy (due to tunneling) nor is it absent at higher energies, for which classical particles escape unimpeded.

Performing a scale transformation as in the main text, we can take $a=\pi$ w.l.o.g., upon which the Wronskian becomes
\beq
  W(\omega)=i\sqrt{\omega}\,\frac{\sin\bigl(\sqrt{\omega{-}V}\pi\bigr)}{\sqrt{\omega{-}V}}
  -\cos\bigl(\sqrt{\omega{-}V}\pi\bigr)\;.
\eeql{W-toy}
In order for this expression to properly account for the decaying boundary condition for $x\To\infty$ and $\omega<0$, one has to define $\sqrt{\omega}$ with the branch cut on the negative imaginary axis; in, e.g., Mathematica, this can be done as $\sqrt{\omega}\equiv e^{i\pi/4}\texttt{Sqrt}[-i\omega]$.\footnote{Since $W(\omega)$ in (\ref{W-toy}) is even in $\sqrt{\omega{-}V}$, there is no branch ambiguity for the latter.} The real roots of $W(\omega)$ can be found graphically as in elementary quantum mechanics; one finds that there are $n$ bound states for $-\frac{1}{4}(2n{+}1)^2<V<-\frac{1}{4}(2n{-}1)^2$. In general, it is convenient to first explore the spectrum by plotting $|W(\omega)|^{-1}$; once the resonances have been located approximately, numerical minimization of $|W|^2$ reliably determines their accurate value. The first few resonances for two instances of the model are given in Table~\ref{toy-tab}. Based on further exploration, it is conjectured that $\re\omega_n>V$ in all cases.

\begin{table}
\begin{tabular}{|c|c|c|c|} \hline
$V$ & $n$ & $\re\omega_n$ & $\im\omega_n$ \\
\hline\hline $-10$ & 1 & $-9.17669$ & 0 \\
\hline $-10$ & 2  & $-6.73671$ & 0 \\
\hline $-10$ & 3 & $-2.82610$ & 0 \\
\hline $-10$ & 4 & 2.01331 & $-0.990207$ \\
\hline $-10$ & 5 & 9.91701 & $-2.52460$ \\
\hline $-10$ & 6 & 19.8311 & $-4.00614$ \\
\hline $-10$ & 7 & 31.7534 & $-5.55510$ \\
\hline $-10$ & 8 & 45.6824 & $-7.18077$ \\
\hline $-10$ & 9 & 61.6167 & $-8.87975$ \\
\hline $-10$ & 10 & 79.5557 & $-10.6465$ \\
\hline $-10$ & 11 & 99.4986 & $-12.4755$ \\
\hline $-10$ & 12 & 121.445 & $-14.3618$ \\
\hline
\end{tabular}\qquad\qquad
\begin{tabular}{|c|c|c|c|} \hline
$V$ & $n$ & $\re\omega_n$ & $\im\omega_n$ \\
\hline\hline 10 & 1 & 10.9716 & $-0.194524$ \\
\hline 10 & 2 & 13.9003 & $-0.748741$ \\
\hline 10 & 3 & 18.8110 & $-1.59648$ \\
\hline 10 & 4 & 25.7197 & $-2.67104$ \\
\hline 10 & 5 & 34.6330 & $-3.92150$ \\
\hline 10 & 6 & 45.5525 & $-5.31221$ \\
\hline 10 & 7 & 58.4781 & $-6.81835$ \\
\hline 10 & 8 & 73.4092 & $-8.42212$ \\
\hline 10 & 9 & 90.3452 & $-10.1103$ \\
\hline 10 & 10 & 109.285 & $-11.8729$ \\
\hline 10 & 11 & 130.229 & $-13.7018$ \\
\hline 10 & 12 & 153.176 & $-15.5908$ \\
\hline
\end{tabular}
\caption{Numerically determined roots of the Wronskian in (\ref{W-toy}), corresponding to the first few resonance frequencies of the square-well model~(\ref{toy}).}
\label{toy-tab}
\end{table}

Clearly, the low-lying bound states for large negative $V$ can be found by approximating the well as a box with hard walls, as $\omega_n\approx n^2+V$. Here, we focus on the less-known high-order modes, with $|\omega|\gg|V|$. For this purpose, set $\nu\equiv\omega-V$ and expand in~$V$. The resonances follow from $e^{2i\pi\sqrt{\nu}}\approx4\nu/V+1$, which is asymptotically solved by
\beq
  \nu_m\sim m^2-i\frac{m}{\pi}\ln\biggl(\frac{4m^2}{V}\biggr)-\frac{\ln^2m}{\pi^2}+\cdots\;.
\eeql{toy-asymp}
For $V<0$, different choices of the logarithm simply yield predictions for different modes. Comparison with Table~\ref{toy-tab} shows good agreement for either choice of $\sgn V$, although the presence of various logarithmic terms in (\ref{toy-asymp}) means that convergence is slow. Interestingly, the dominant term $m^2$ is the same as for an infinite well. For $V\To0$, the decay rates diverge as expected, but perhaps counterintuitively, for any fixed $V$ the quality factor $Q_m=\re\omega_m/2\lvert\im\omega_m\rvert$ increases without bound.

For comparison, we briefly mention the model $U(x)=K\delta(x-\pi)$, for which the modes are asymptotically
\beq
  \omega_n\sim n^2-i\frac{n}{\pi}\ln\biggl(-\frac{2in}{K}\biggr)
  -\frac{\ln^2n}{4\pi^2}+\cdots\;.
\eeq
Comparing with (\ref{toy-asymp}), one sees that the stronger potential singularity ($\delta$-peak versus step) causes stronger scattering, and hence a decay rate which asymptotically is a factor two smaller.

A closed-form Wronskian (in terms of hypergeometric and related functions) should also be obtainable for many models with a potential consisting of two constant, linear, or (inverted-)harmonic pieces, on either the full or the positive $x$-axis. However, unlike any of these, the cubic potential is infinitely smooth, so that its asymptotic scattering properties (reflected in the damping rates of high-order resonances) are expected to be qualitatively different.

\end{document}